\documentclass[12pt]{iopart}

\usepackage{graphicx}
\usepackage{amssymb}
\usepackage{epstopdf}
\usepackage{graphicx}
\usepackage{amssymb}
\usepackage{amsthm}
\usepackage{harvard}
\usepackage{tikz}
\usetikzlibrary{arrows,decorations.pathmorphing,backgrounds,positioning,fit,petri,patterns}
\usepgflibrary{patterns} 
\usepackage[nooneline,scriptsize]{subfigure}

\newcommand{\degree}{\ensuremath{^\circ}}

\usepackage{iopams}

\begin{document}

\title{Optimal partial-arcs in VMAT treatment planning}

\vspace{2pc}


\author{Jeremiah Wala, Ehsan Salari, Wei Chen and David Craft}

\address{Department of Radiation Oncology, Massachusetts General Hospital, 30 Fruit Street, Boston, MA 02114, USA}

\ead{jeremiah\_wala@hms.harvard.edu}

\begin{abstract}

\noindent \textbf{Purpose}: To improve the delivery efficiency of VMAT by extending the recently published VMAT treatment planning algorithm {\sc vmerge} to automatically generate optimal partial-arc plans.

\noindent\textbf{Methods and materials}: A high-quality initial plan is created by solving a convex multicriteria optimization problem using 180 equi-spaced beams. This initial plan is used to form a set of dose constraints, and a set of partial-arc plans is created by searching the space of all possible partial-arc plans that satisfy these constraints. For each partial-arc, an iterative fluence map merging and sequencing algorithm ({\sc vmerge}) is used to improve the delivery efficiency. Merging continues as long as the dose quality is maintained above a user-defined threshold. The final plan is selected as the partial-arc with the lowest treatment time. The complete algorithm is called {\sc pmerge}. 

\noindent\textbf{Results}: Partial-arc plans are created using {\sc pmerge} for a lung, liver and prostate case, with final treatment times of 127, 245 and 147 seconds. Treatment times using full arcs with {\sc vmerge} are 211, 357 and 178 seconds. Dose quality is maintained across the initial, {\sc vmerge}, and {\sc pmerge} plans to within 5\% of the mean doses to the critical organs-at-risk and with target coverage above 98\%. Additionally, we find that the angular distribution of fluence in the initial plans is predictive of the start and end angles of the optimal partial-arc.

\noindent\textbf{Conclusions}: The {\sc pmerge} algorithm is an extension to {\sc vmerge} that automatically finds the partial-arc plan that minimizes the treatment time. VMAT delivery efficiency can be improved by employing partial-arcs without compromising dose quality. Partial-arcs are most applicable to cases with non-centralized targets, where the time savings is greatest.

 \end{abstract}

\submitto{Physics in Medicine and Biology}

\maketitle

\section{Introduction}

Volumetric modulated arc therapy (VMAT) is an extension to intensity modulated radiation therapy (IMRT) that offers the ability to deliver highly conformal dose distributions in a fraction of the time as IMRT~\cite{Yu2011}. In both techniques, the delivered radiation is modulated by a multi-leaf collimator (MLC) that conforms the dose to the targets. However, unlike IMRT where radiation is delivered at a fixed set of beam angles, in arc therapy the gantry delivers radiation continuously as it moves in an arc around the patient. In VMAT, one of the more widely used forms of arc therapy, the gantry speed, dose rate, and MLC leaf positions are allowed to vary as the gantry moves around the couch~\cite{Otto2008}. The additional degrees of freedom afforded by VMAT have been shown to maintain or improve the dose conformity when compared with IMRT~\cite{Teoh2011,Palma2008,Cozzi2008,Johnston2011}. However, this also makes the treatment planning process much more complex. 

One of the most compelling reasons for using VMAT over IMRT is the ability to reduce the treatment time, often below 5 minutes per session~\cite{Bedford2009,Clivio2009,Rao2010,Zhang2010b}. Reducing treatment time and the number of monitor units delivered reduces errors from patient motion~\cite{Hoogeman2008}, decreases the scattered dose responsible for secondary malignancies~\cite{Hall2003,Ruben2008,Kry2005}, and may have direct radiobiological benefits~\cite{Wang2003,Bese2007}. In addition, it reduces the amount of time patients need to spend receiving their therapy, and may allow for the treatment of more patients. 

Approaches in reducing treatment time have focused on, for example, optimizing the delivery parameters (e.g. dose rate) given an initial plan~\cite{Rangaraj2010,Boylan2011}, improving the delivery control system~\cite{Bertelsen2011} and developing leaf sequencing algorithms that improve delivery efficiency~\cite{Luan2008}. In our own work on VMAT, we introduced {\sc vmerge}, a method to facilitate the selection of the best balance between planning objectives and delivery efficiency, one of the main tradeoffs in VMAT treatment planning~\cite{Craft2012}. Inherent to most approaches is the idea that in the tradeoff between dose quality and delivery efficiency, dose quality should be highly favored. 

Arc therapy can be delivered as a partial-arc, a full single arc, or as multiple arcs. For complex cases where a large amount of beam modulation is required, delivery in a single arc requires significant slowing of the gantry and reduction of the dose rate to achieve dose distributions similar to IMRT~\cite{Bortfeld2009}. Instead, radiation may be delivered over multiple arcs, which provides the necessary degrees of freedom but will tend to increase the total treatment time and required number of monitor units~\cite{Guckenberger2009}. Alternatively, for cases with peripherally located targets, one might expect that appreciable beam modulation is only required over a critical angle range. In this case, the gantry might move at maximal speed and with small aperture sizes over the angles that do not provide for effective radiation. Eliminating this portion of the arc from the delivery would be expected to reduce the treatment time with minimal effect on the dose distribution. However, to the best of our knowledge, no VMAT algorithm explicitly addresses this situation. In this work, we present a method called {\sc pmerge} for automatically finding the optimal partial-arcs that eliminate the unneeded arc segments without significantly compromising dose quality.

Like the beam angle optimization problem for fixed beam IMRT, the partial-arc optimization
problem is nonconvex. Nonconvexity means that a local minimum is not guaranteed to be a 
global minimum. In order to guarantee a global minimum for nonconvex problems,
a global search needs to be performed. 
Given that the nonconvex space of the partial-arc optimization 
problem consists of only two variables, the start and end gantry angles, this global search 
can be done by discretizing the search space and enumerating all possible start/end angle pairs, which 
is the strategy we adopt.  This strategy is inherently parallelizeable, allowing for shorter computation
times.


To solve the VMAT problem for each partial-arc we use our previously reported merging and sequencing algorithm called {\sc vmerge}~\cite{Craft2012}. We call the complete process {\sc pmerge}, which employs both merging and sequencing and an enumerable set of partial-arcs to reduce treatment time. Given an initial high quality plan as a dose ``gold standard'', {\sc pmerge} will automatically determine the fluence maps, dose rate, gantry speed, and starting and ending gantry angles that minimize the treatment time while maintaining the dose distribution to within a user-defined quality threshold. 

\section{Materials and methods} \label{sec:method}


We define the following terms: \emph{initial plan}, \emph{partial-arc plan}, \emph{planning goals},  {\sc vmerge} plan and {\sc pmerge} \emph{plan}. The \emph{initial plan} refers to a high-quality plan that serves as the dose standard used to constrain all partial-arcs. The initial plan is created without regards to delivery efficiency, providing a standard for understanding the tradeoff between dose quality and delivery efficiency. Our strategy for creating the initial plan is described in Section~\ref{sec:initialopt}, where we define a 180-beam IMRT problem, with beams placed every $2\degree$, and solve it using multi-criteria optimization (MCO). 

\begin{figure}[h] 
\centerline{\includegraphics[width=5in]{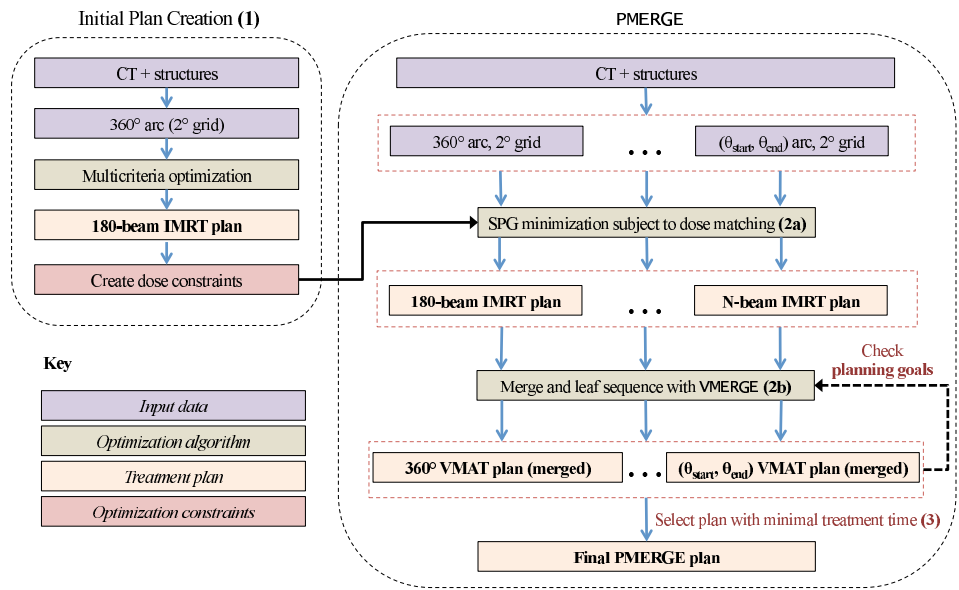}}
\caption{Illustration of our three step process. 1) An initial IMRT plan is created to generate a set of dose constraints for {\sc pmerge}. 2a) {\sc pmerge} uses these constraints to generate a set of partial-arc solutions. 2b) Each $N$-beam partial-arc solution is merged and sequenced with {\sc vmerge}, subject to the user-defined planning goals. 3) The final plan is selected as the one with the lowest treatment time.}
\label{fig:chart}
\end{figure}

Next, a set of \emph{partial-arc plans} is created using a two-step process. First, we solve a set of restricted angle IMRT problems with various start and end gantry angles, with beams placed every $2\degree$, by finding feasible solutions that minimize the sum-of-positive gradients (SPG), subject to matching the initial dose quality (Section~\ref{sec:partarc}). We then use {\sc vmerge} to iteratively merge neighboring fluence maps to create deliverable VMAT plans with improved delivery efficiency. Merging is automatically halted when the merged plan violates a set of user-defined \emph{planning goals}, which articulates how much the planner is willing to deviate from the initial plan in order to improve delivery efficiency. We refer to the {\sc vmerge} plan as the result of the merging algorithm on the 180-beam plan, and the {\sc pmerge} plan as the merged partial-arc with the lowest treatment time. The full method is outlined in Figure~\ref{fig:chart}.

\subsection{Delivery parameters and leaf sequencing} \label{sec:delparam}

There are two classes of methods for solving VMAT problems: one-stage and two-stage~\cite{Yu2011}. In the one stage method, known as \emph{direct aperture optimization}, the paths of the MLC leaves are directly optimized, taking into account limitations of finite leaf speed.  Instead, we use a two-stage planning approach to solve the VMAT problem for any size arc. In the first step, an $N$-beam IMRT problem with beams spaced every $2\degree$ is solved to generate a set of $N$ fluence maps ($f^{1}_{l,x}$, $f^2_{l,x}$..., $f^{N}_{l,x}$), where each fluence map is a K$\times$L matrix of $L$ rows of MLC leaves and $K$ columns. In the second step, a leaf sequencing algorithm determines the MLC leaf trajectories, gantry speed ($\omega$), and dose rate ($r$) for each 2$\degree$ arc portion, which we call a {\it gantry sector}.  In leaf sequencing, the trajectories of the leaves are set so that the accumulated fluence over a given gantry sector matches the input fluence from the static beam fluence map. To accomplish this, we employ a widely used dynamic MLC (dMLC) sequencer which creates a ``sliding window'' that moves leaves unidirectionally across the field for each arc portion~\cite{Spirou1994,Svensson1994}. Thus, for each gantry sector, the leaves must move from left-to-right (or right-to-left), traversing the entire field of width $W$. In practice, the gantry speed, dose rate and maximum leaf speed are constrained by the limitations of the linac hardware and will have a significant effect on the total treatment time~\cite{Boylan2011}. We select a set of hardware capabilities (Table~\ref{tab:constraints}) that are representative of the values found on many current treatment machines.   

\begin{table}[h]\footnotesize \label{tab:constraints}
\begin{center}

\caption{\label{tab:constraints}VMAT delivery parameters and constraints}
\begin{indented}
\begin{tabular}{lll}
\br
Delivery parameter & Symbol & Constraint \\
\hline \hline
Maximum leaf speed & $v_{\max}$ & 2.5 cm/s \\
Maximum gantry speed & $\omega_{\max}$ & 6 deg/s \\
Minimum gantry speed &$\omega_{\min}$ & 0 deg/s \\
Maximum dose rate & $r_{\max} $ & 600 MU/min \\ 
Minimum dose rate & $r_{\min} $ & 50 MU/min\\
\br
\end{tabular}
\end{indented}
\end{center}
\end{table}

When applied to VMAT, the total treatment time using the dMLC algorithm is controlled by three terms: 1) the total distance the leaves must travel during delivery, 2) the amount of beam modulation required to deliver the desired fluence map and 3) the maximum gantry speed. These three considerations can be combined to provide an equation to calculate the total treatment time for a plan with $N$ gantry sectors of arc lengths $\Delta\theta_i$:

\begin{equation}
\label{eqn:tt}
T = \displaystyle\sum_{i=1}^N\max\left[\frac{W}{v_{\max}} +\frac{\displaystyle\max_{\mbox{rows$\;l$}} \left[\displaystyle\sum_{x=1}^K\left(\frac{\mbox{d}f_{l,x}^i(x)}{\mbox{d}x}\right)^+\right]}{r_i},\frac{\Delta\theta_i}{\omega_{\max}}\right]
\end{equation}

\noindent where the (.)$^+$ operator is shorthand for $\max\left(0,.\right)$. The first term in (\ref{eqn:tt}) is the field width divided by the maximum leaf speed, which is the minimum time required for the leaves to move from the left side of the field to the right (or vice versa) if no fluence were delivered. As fluence is added, the leaf speed will be modulated to allow enough time for the dose to accumulate. This time is described in the second term as the sum-of-positive gradients (SPG), a measure of the amount of variation in a field~\cite{Craft2007a}, divided by the dose rate. Fields that are highly varied (have a high SPG term) will take longer to deliver with dMLC. Finally, because of limitations on the maximum gantry speed, the time required to deliver a gantry sector cannot be less than time required to move the gantry at maximum speed over the gantry sector. The total treatment time is the sum of the time required to deliver each of the $N$ gantry sectors.

Given a fixed set of hardware capabilities, the total treatment time can be improved by either decreasing the total number of gantry sectors $N$ or by decreasing the SPG. Decreasing the number of gantry sectors motivates the partial-arc strategy described here, and the merging strategy {\sc vmerge} -- by using partial-arc solutions, we begin with a lower number of fluence maps than in a full arc, and this is decreased further by combing gantry sectors in the merging routine. Thus, for each gantry sector that is merged or eliminated with a partial-arc, the treatment time decreases by no less than $W/v_{\max}$. Finally, to further reduce treatment time, we incorporate SPG minimization into our optimization step as described in Section~\ref{sec:partarc}. 


\subsection{Initial plan optimization} \label{sec:initialopt}

We use a multicriteria optimization (MCO) approach to solve the initial 180-beam IMRT problem, as described in~\citeasnoun{Craft2012}. Briefly, in MCO, each structure is assigned an objective function, and a tradeoff surface of Pareto optimal plans is generated by the optimizer~\cite{Monz2008,Thieke2007}. A Pareto optimal plan is one where no improvements could be made to any objective without worsening another objective~\cite{cotrutz2001}. The user navigates this tradeoff surface and selects the plan that best meets the treatment goals. 

The general multicriteria IMRT formulation is the following:

\begin{eqnarray}
\hbox{optimize~~~} & \{g_1(d),~g_2(d), \ldots g_M(d)\} \nonumber\\
\hbox{subject~to~~~}   & d = Df \nonumber\\
~~~~ & d \in C \nonumber\\
~ &  f \ge 0
\label{mco}
\end{eqnarray}

\noindent where $g_i$ are the $M$ objective functions describing the dose to a particular structure (e.g., mean dose to stomach), $d$ is the vector of voxel doses, $D$ is the dose-influence matrix, and $f$ is a concatenation of all the fluence maps into a single beamlet fluence vector. The set $C$ is a convex set of dose constraints (e.g. maximum dose to the target) that are designed to be met by all plans. We use the CERR 3.0~\cite{deasy2003} environment and its beamlet based quadratic infinite beam (QIB) dose computation to calculate the dose influence matrix, with a computation time of around 10 minutes per case. Given a dose matrix, the dose distribution can be calculated from the fluence maps in seconds.

The constraints on the planning target volumes (PTV) and the OARs for each site are listed in Table~\ref{tab:MCO}. In our MCO formulation, we had the optimizer minimize the mean doses to the OARs. Additionally, we include as an OAR unclassified tissue (u.t.), which encompasses all voxels that do not belong to another structure. Optimizations are performed using the solver described by~\citeasnoun{Chen2010}. We then choose a plan for each treatment site by exploring the Pareto surface and selecting a plan that balanced the dose the OARs.

\begin{table}[h]
\begin{center}
\caption{ \label{tab:MCO}Target coverage constraints and OAR list used in the MCO optimization}
\begin{tabular}{lll}
\\
{\bf Treatment site} & {\bf PTV Min; Max (Gy)} & {\bf OARs}\\
\hline\hline
Lung& 50.0; 55.0 & left lung, right lung, esophagus,\\
& & heart, spinal cord, u.t. \\
\hline 
Liver& 50.0; 55.0 & left lung, right lung, liver, spinal cord, \\
& & stomach, left kidney, right kidney, u.t.\\
\hline 
Prostate& 79; 85.3 & left femoral head, right femoral head, \\
& & anterior rectum, bladder, u.t. \\
\hline 
\end{tabular}
\end{center}
\end{table}

\subsection{Generating the partial-arc plans} \label{sec:partarc}

The set of partial-arc solutions are created by iteratively solving the $2\degree$ grid IMRT problem for a range of partial-arcs. In creating the plans, we respect the constraint that the gantry cannot rotate through $0\degree$, a line from the couch to the floor. Plans are created with arc lengths every 40$\degree$ ($360\degree, 320\degree$, ...) until there are no feasible solutions. For each arc length, plans are created with arcs centered at intervals of $40\degree$. Thus, there is one $360\degree$ plan, three $320\degree$ degree plans (beginning at $-40\degree, 0\degree, 40\degree$), five $280\degree$ plans, etc.

Each partial-arc solution is created to maintain the same dose quality as the initial 180-beam solution. This is done with the same solver used to solve the original 180-beam MCO formulation, except run in feasibility mode, where the solver returns the first solution it finds which satisfies all of the hard constraints. Because treatment time scales with the complexity of the solution, this solution is then smoothed using total SPG as the objective function, while maintaining dosimetric feasibility: \\ \\
\noindent {\bf SPG minimization subject to dose matching}
\begin{eqnarray}
\hbox{minimize~~~} & \hbox{$\displaystyle\sum_{i=1}^N\max_{\mbox{rows$\;l$}} \left[\displaystyle\sum_{x=1}^K\left(\frac{\mbox{d}f_{l,x}^i(x)}{\mbox{d}x}\right)^+\right]$} \nonumber\\
\hbox{subject~to~~~}   & d = Df \nonumber\\
~~~~ & d_i \ge B_L ,~ \forall i \in \hbox{~target}\nonumber\\
~~~~ & d_i \le B_U ,~ \forall i \nonumber\\
~~~~ & \mbox{mean}_j^{(P)}(d)\le \hbox{mean}^{(I)}_j(d),~ \forall  \hbox{~OAR}\;j\nonumber\\
~ &   f \ge 0
\label{eqn:spg}
\end{eqnarray}

\noindent where $B_L$ and $B_U$ is the target lower bound, $B_U$ is the upper bound on the dose,  and $\hbox{mean}^{(P)}_j(d)$ and mean$_j^{(I)}(d)$ are the mean doses to the $j$th OAR for the partial-arc and initial plans, respectively. This formulation ensures that for each partial-arc solution the target coverage is no less than, and the maximum dose and the mean doses to each OAR are no greater than those of the initial plan. 

\subsection{Merging and final plan selection} \label{sec:merge}

After the SPG minimization step, we have a set of IMRT solutions for various partial-arcs. To improve the delivery efficiency, we use a previously described fluence map merging strategy called {\sc vmerge} to combine gantry sectors~\cite{Craft2012}. Briefly, {\sc vmerge} employs a greedy merging strategy to iteratively add similar neighboring fluence maps -- for each iteration, the two neighboring fluence maps that are most similar are added together into a single fluence map. This reduces the number of gantry sectors and improves the delivery efficiency, at the expense of reducing the amount of fluence modulation and the angular resolution.

{\sc pmerge} automates {\sc vmerge} by utilizing a quality cutoff, which limits how much the dose can change, and automatically stops merging when the quality drops below the planning goals. Here, the planner is free to chose how much the dose quality is allowed to change in order to improve delivery efficiency. We use the initial high-quality plan as a guide for how to set these goals. For the OARs that receive the most radiation, we limit the partial-arc mean dose to less than 105\% of the initial mean dose for that OAR. Additionally, we ensure that the target coverage stays above 98\%, and the maximum dose is less than 102\% of the initial maximum dose. Finally, we illustrate the use of site-specific planning goals, including limiting the maximum dose to the femoral heads in the prostate case and adding additional constraints on the left lung dose in the lung case. The full sets of planning goals are provided in Tables 3-5.

\section{Results}

\subsection{Lung}

\begin{table}[h]
\begin{center}
\caption{\label{tab:lung}\textbf{Lung}: Planning goals, dose quality and treatment times}
\begin{tabular}{llll}
\br
\textbf{}&\textbf{Goal}&\textbf{Initial (I)}& {\sc pmerge} \textbf{(P)}\\
\hline
 u.t. dose & $\hbox{mean}^{(P)}_{UT} < 1.10\times \hbox{mean}^{(I)}_{UT}$ & 2.8 Gy & 3.0 Gy\\
Right lung & $\hbox{mean}^{(P)}_{RL} < 1.05\times \hbox{mean}^{(I)}_{RL}$&0.6 Gy & 0.5 Gy \\
Left lung  &$^\dag V40^{(P)}<1.05\times V40^{(I)}$&9.4\% & 9.5\% \\
&$V50^{(P)}<1.05\times V50^{(I)}$&3.0\%&3.0\%\\
Target coverage &$V50^{(P)} > 0.98\times V50^{(I)}$&100\% & 98.2\% \\ 
Maximum dose &$\max^{(P)} < 1.02\times \max^{(I)}$&55.0 Gy & 55.8 Gy\\
\hline
(Start, end) angles && (0$\degree$, 360$\degree$) & (60$\degree$, 260$\degree$) \\
Treatment time$^\ddag$ && 211 s &127 s\\
\br
\multicolumn{4}{l}{$\dag$V$x$ denotes the volume receiving at least $x$ Gy} \\
\multicolumn{4}{l}{$\ddag$Treatment times are after merging} \\
\end{tabular}
\end{center}
\end{table}

\begin{figure}[H] 
\centerline{\includegraphics[width=6in]{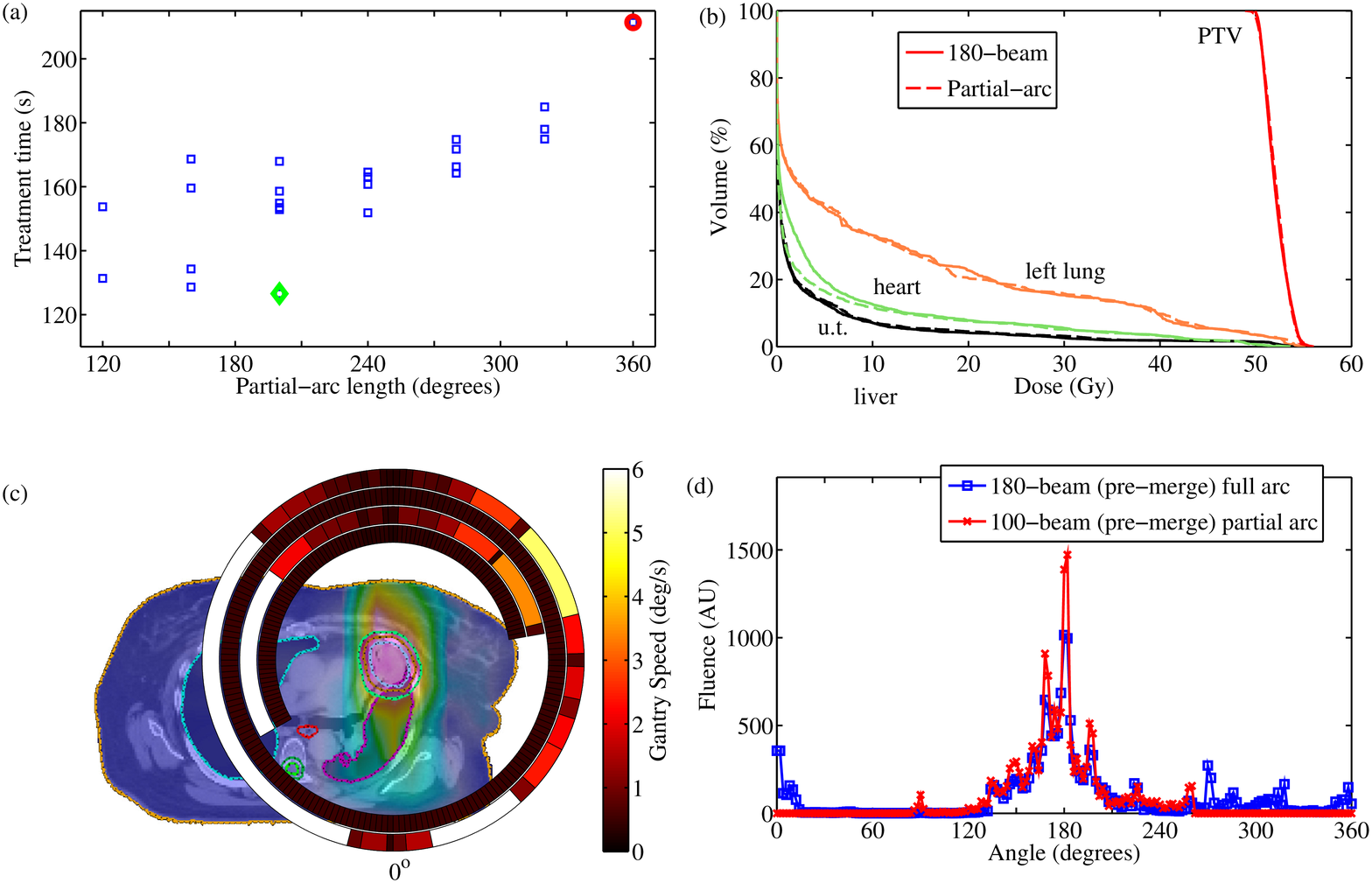}}
\caption{\textbf{Lung}: a) Final treatment times for each of the merged partial-arc plans (blue squares) with treatment times near or below the {\sc vmerge} plan (red circle). b) DVH for the initial 180-beam plan (solid) and the {\sc pmerge} plan (dashed). c) Initial and merged gantry sectors for the {\sc pmerge} plan (inner rings) and 180-beam plan (outer rings) over the {\sc pmerge} dose wash. d) Distribution of fluence for the initial solution and the optimal partial-arc before merging  (100-beam). }
\label{fig:lung}
\end{figure}

We navigate to an initial 180-beam plan with mean doses and target coverage as shown in Table~\ref{tab:lung}. The treatment time of the initial plan before merging is 771 s, and is reduced to 211 s with {\sc vmerge}. The initial plan and {\sc vmerge} plans are shown with the outer two rings of Fig~\ref{fig:lung}(c). The {\sc vmerge} plan is notable for a large arc portion of $116\degree$ on the contralateral side of the PTV where the gantry speed is at its maximum. As shown in Figure~\ref{fig:lung}(d), there is little fluence delivered over this section in the initial 180-beam plan.

Partial-arc solutions were attempted for arcs with lengths between $80\degree$ and $360\degree$, at intervals of $40\degree$. 50 had feasible solutions that met the dose constraints from Equation~\ref{eqn:spg}, and 26 of these had lower treatment times than the {\sc vmerge} plan. The final {\sc pmerge} plan has an arc length of $200^{\circ}$ and treatment time of 127 s, and is shown in the inner two rings of Fig 2c, both before and after merging. The total difference in treatment times between the {\sc vmerge} (211 s) and {\sc pmerge} (127 s) plans is 84 s, a 40\% reduction.  Optimization, merging and sequencing took about 40 minutes to complete all 64 plans, running in parallel on 8 processors. Computation time was similar across the lung, liver and prostate cases. 

\subsection{Liver}

\begin{table}[h]
\begin{center}
\caption{\label{tab:liver}\textbf{Liver}: Planning goals, dose quality and treatment times}
\begin{tabular}{llll}
\br
\textbf{}&\textbf{Goal}&\textbf{Initial (I)}& {\sc pmerge} \textbf{(P)}\\
\hline 
u.t.  &$\hbox{mean}^{(P)}_{UT} < 1.10\times \hbox{mean}^{(I)}_{UT}$& 5.7 Gy & 5.6 Gy\\
Right lung  &$\hbox{mean}^{(P)}_{RL} < 1.05\times \hbox{mean}^{(I)}_{RL}$& 9.1 Gy & 9.2 Gy \\
Left lung  & no constraint (mean$_{LL}^{(P)})$& 1.5 & 1.5 Gy \\
Liver  &$\hbox{mean}^{(P)}_{L} < 1.05\times \hbox{mean}^{(I)}_{L}$& 23.3 Gy & 23.3 Gy \\
Right kidney  &$\hbox{mean}^{(P)}_{RK} < 1.05\times \hbox{mean}^{(I)}_{RK}$& 5.8 Gy & 4.7 Gy \\
Left kidney  & no constraint (mean$_{LK}^{(P)})$& 0.7 Gy & 0.4 Gy \\
Target coverage &$V50^{(P)} > 0.98\times V50^{(I)}$&100\% & 98.0\% \\ 
Maximum dose &$\max^{(P)} < 1.02\times \max^{(I)}$&55.0 Gy & 55.6 Gy\\
\hline
(Start, end) angles && (0$\degree$, 360$\degree$) & (100$\degree$, 340$\degree$) \\
Treatment time$^\dag$ && 357 s & 245 s\\
\br
\multicolumn{4}{l}{$\dag$Treatment times are after merging} \\
\end{tabular}
\end{center}
\end{table}

\begin{figure}[H] 
\centerline{\includegraphics[width=6in]{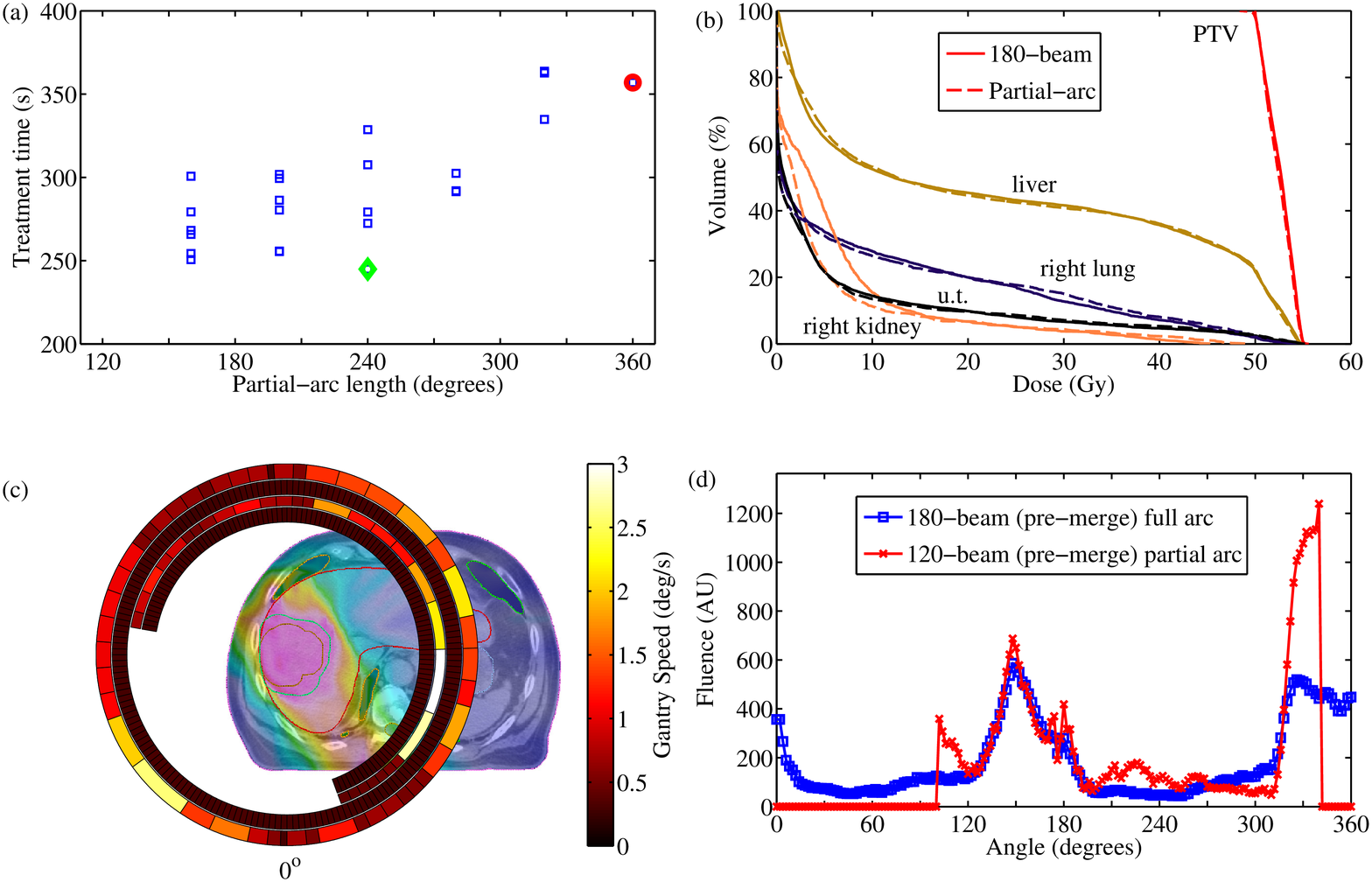}}
\caption{\textbf{Liver}: a) Final treatment times for each of the merged partial-arc plans (blue squares) with treatment times near or below the {\sc vmerge} plan (red circle). b) DVH for the initial 180-beam plan (solid) and the {\sc pmerge} plan (dashed). c) Initial and merged gantry sectors for the {\sc pmerge} plan (inner rings) and 180-beam plan (outer rings) over the {\sc pmerge} dose wash. d) Distribution of fluence for the initial solution and the optimal partial-arc before merging (120-beam).  }
\label{fig:liver}
\end{figure}

We navigate to an original 180-beam solution with mean OAR doses, target coverage and maximum dose as shown in Table~\ref{tab:liver}. The treatment time before merging for the full arc solution is 1214 s and is reduced to 357 s with {\sc vmerge} (Figure~\ref{fig:liver}(c), outer rings). Partial-arc solutions were attempted for arcs lengths between $80\degree$ and $360\degree$, at intervals of $40\degree$, and 33 had feasible solutions. There were no solutions with arc lengths less than $160\degree$. Of the feasible partial-arcs, 19 had lower treatment times than the {\sc vmerge} plan.

The selected partial-arc has an arc length of $240^{\circ}$ (Figure~\ref{fig:liver}(c), inner rings) and treatment time of 245 s. The total difference in treatment times between the {\sc vmerge} plan (357 s) and {\sc pmerge} plan (245 s) is 112 s, a 31\% reduction. In this case the initial 180-beam solution is significant for two peaks of fluence, one at approximately $150\degree$ and one at $330\degree$, as shown in Figure~\ref{fig:liver}(d). The optimal partial-arc includes both of these peaks, and is able to eliminate the first $100\degree$ where the amount of required fluence is low.

\subsection{Prostate}

\begin{table}[h]
\begin{center}
\caption{\label{tab:prostate}\textbf{Prostate}: Planning goals, dose quality and treatment times}
\begin{tabular}{llll}
\br
\textbf{}&\textbf{Goal}&\textbf{Initial (I)}& {\sc pmerge} \textbf{(P)}\\
\hline 
u.t.  &$\hbox{mean}^{(P)}_{UT} < 1.10\times \hbox{mean}^{(I)}_{UT}$& 15.5 Gy & 16.4 Gy\\
Anterior rectum  &$\hbox{mean}^{(P)}_{AR} < 1.05\times \hbox{mean}^{(I)}_{AR}$& 41.4 Gy & 41.8 Gy \\
Bladder &$\hbox{mean}^{(P)}_{B} < 1.05\times \hbox{mean}^{(I)}_{B}$& 39.6 Gy & 39.8 Gy \\
Left femoral head &$\max^{(P)}_{LFH} < 1.10\times \max^{(I)}_{LFH}$& 33.6 Gy & 35.5 Gy \\
Right femoral head&$\max^{(P)}_{RFH} < 1.10\times \max^{(I)}_{RFH}$& 37.0 Gy & 38.3 Gy \\
Target coverage &$V79^{(P)} > 0.98\times V79^{(I)}$&100\% & 98.2\% \\ 
Maximum dose &$\max^{(P)} < 1.02\times \max^{(I)}$&85.3 Gy & 86.7 Gy\\
\hline
(Start, end) angles && (0$\degree$, 360$\degree$) & (40$\degree$, 320$\degree$) \\
Treatment time$^\dag$ && 178 s & 147 s\\
\br
\multicolumn{4}{l}{$\dag$Treatment times are after merging} \\
\end{tabular}
\end{center}
\end{table}

\begin{figure}[H] 
\centerline{\includegraphics[width=6in]{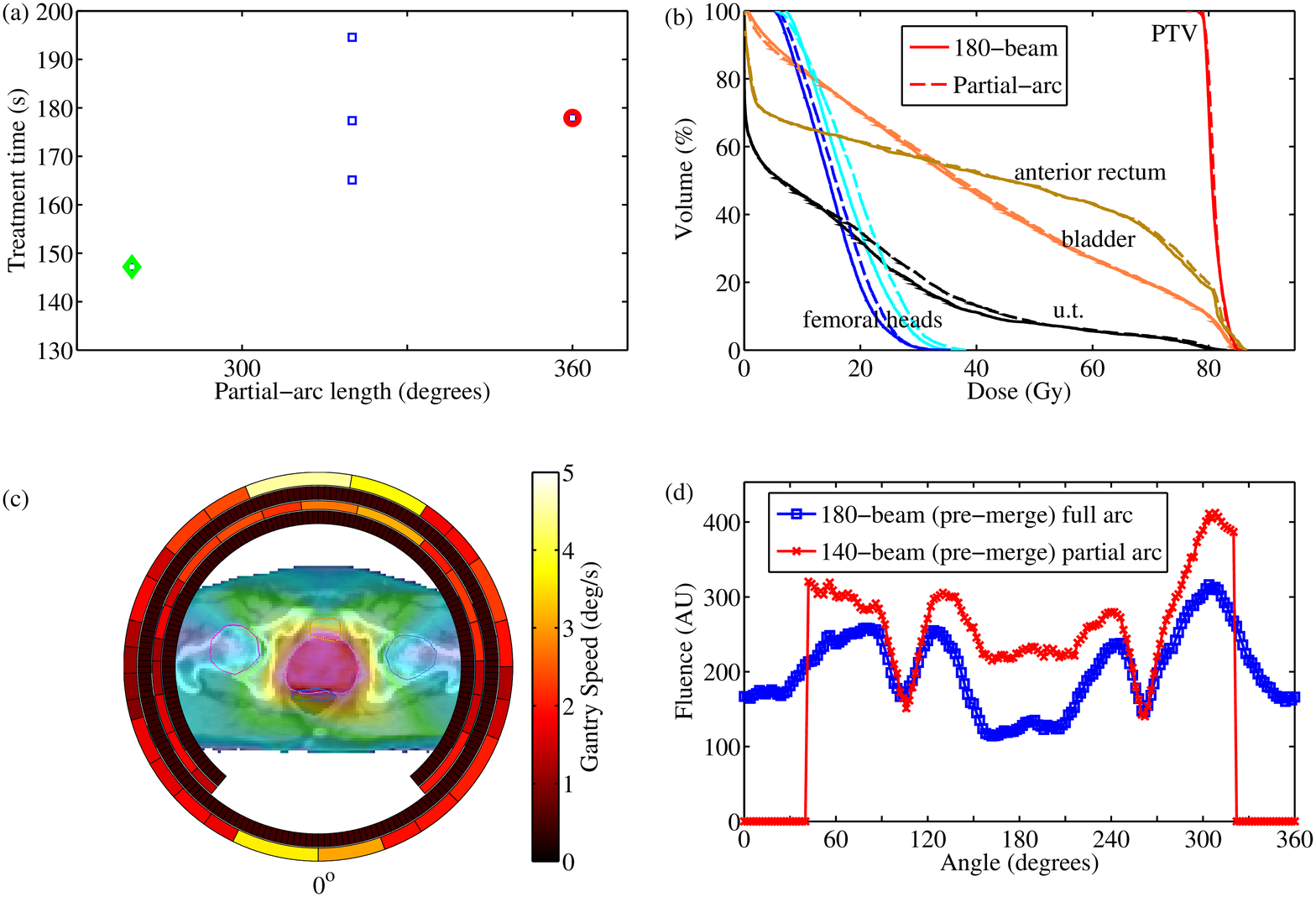}}
\caption{\textbf{Prostate}: a) Final treatment times for each of the merged partial-arc plans (blue squares) with treatment times near or below the {\sc vmerge} plan (red circle). b) DVH for the initial 180-beam plan (solid) and the {\sc pmerge} plan (dashed). c) Initial and merged gantry sectors for the {\sc pmerge} plan (inner rings) and 180-beam plan (outer rings) over the {\sc pmerge} dose wash. d) Distribution of fluence for the initial solution and the optimal partial-arc before merging (140-beam).  }
\label{fig:prostate}
\end{figure}

We navigated to an original 180-beam solution with mean OAR doses, target coverage and maximum dose as shown in Table~\ref{tab:prostate}. The treatment time before merging for the full arc solution is 860 s and is reduced to 178 s with {\sc vmerge} (Figure~\ref{fig:prostate}(c), outer two rings). Unlike the lung and liver cases, the {\sc vmerge} plan has regularly spaced arc portions of similar size, and the fluence is delivered fairly uniformly around the target, as shown in Figure~\ref{fig:prostate}(d). This is reflected in the optimal partial-arc solution, which is highly similar to the initial plan through nearly the full arc. 

Partial-arc solutions were created for arcs lengths between $80\degree$ and $360\degree$, at intervals of $40\degree$, and 45 had feasible solutions. Of the 45, only 3 had treatment times lower than the {\sc vmerge} plan. There were no feasible partial-arcs at $80\degree$. For plans with arc lengths of less than 280$\degree$, almost no merging was possible before the plan quality degraded beyond the stopping criteria. The selected partial-arc has an arc length of $300^{\circ}$ (Figure~\ref{fig:prostate}(c), inner rings) and treatment time of 147 s. The total difference in treatment times between the {\sc vmerge} (178 s) plan and {\sc pmerge} plan (147 s) is 31 s, a 17\% reduction. 

\section{Discussion and conclusions}

We present a method for automatically creating efficient VMAT partial-arc plans from a high quality reference plan. By sampling a set of partial-arcs, we reduced treatment times for a lung, liver and prostate case. The final {\sc pmerge} plans reflect the underlying geometry of the problem. For the lung and liver cases with non-centralized PTVs, a signifiant portion of the full-arc can be eliminated with little effect on the dose quality. This allowed for the treatment time to be reduced by 40\% and 31\% respectively. The reduction in treatment time for the prostate case was more modest at 17\%. As might be expected for a centrally located target, the fluence was distributed more evenly around the arc, and most of the partial-arcs were not able to improve the delivery efficiency while maintaining the dose quality to within the specified boundaries.


In {\sc pmerge}, the dose quality of the final partial-arc is designed to be highly similar to the initial plan. However, the {\sc pmerge} plan is limited only by the planning goals specified after the initial plan creation. Thus, the partial-arc plan could deviate from the initial plan in ways that are not specified by this quality threshold. Any number or type of constraints could easily be imposed on the final solution to ensure that the initial plan and {\sc pmerge} plan were sufficiently similar.  In practice, we did not find that it was necessary to impose many constraints in order to match the solutions. Although we were free to use any number of dose-volume points or EUD functions, our planning goals were restricted to only one or two constraints per OAR, usually a limit on the mean or maximum dose. For all three cases, these were sufficient to closely match the doses across the DVH, even for OARs which were not explicitly specified.  

In VMAT, as in IMRT, there are many different ways to deliver the fluence that will give highly similar dose distributions. Because the treatment time is dependent on the geometry of these fluence maps, the method by which the initial fluence maps are created is an important determinant of treatment time.  Although {\sc pmerge} is not confined to any single optimization strategy, we found that using SPG smoothing was critically important to achieving highly efficient plans. Indeed, fluence map smoothing is important in IMRT  (e.g. ~\cite{Matuszak2008}), where it is know that the complexity of the fluence maps -- and hence the number of monitor units -- can be greatly reduced without significantly impacting the dose distribution.

Although the partial-arc global search can be parallelized, we recognize that the high computation time is a limit to our method. The observation that the optimizer distributes fluence very similarly between the full-arc and partial-arc plans suggests a way of approximating the appropriate partial-arc without having to enumerate all of the possibilities. One strategy would be to plot the arrangement of fluence around the arc and have the planner select the partial-arc that captures the critical angles. For the lung and liver cases, we find that there were many different partial-arcs that greatly improved the treatment time, and thus it is not essential to find the absolute global minimum when plans near the minimum are almost equivalent (see Figures~\ref{fig:lung}(a) and~\ref{fig:liver}(a)). More importantly, we conclude that partial-arcs that cover certain critical angles are useful for reducing treatment time in VMAT, and {\sc pmerge} is a useful first-pass method at finding the optimal plan.

\section*{Acknowledgments}
The project described was supported by Award Number R01CA103904 from the National Cancer Institute. The content is solely the responsibility of the authors and 
does not necessarily represent the official views of the National Cancer Institute 
or the National Institutes of Health. The authors would like to thank Dualta McQuaid for his contributions to the computational infrastructure used in this research.
\section*{References}

\bibliographystyle{jphysicsB}
\bibliography{Wala_bibliography}

\end{document}